\documentclass[runningheads]{llncs}
\usepackage{xcolor}
\usepackage[T1]{fontenc}
\usepackage{comment}
\usepackage{algorithm}
\usepackage{algpseudocode}
\usepackage{mathtools}

\algnewcommand{\LineComment}[1]{\State \(\triangleright\) #1}
\usepackage{amsmath}

\usepackage{subcaption}
\usepackage[misc]{ifsym}
\usepackage{graphicx}

\usepackage{hyperref}
\usepackage{amsmath} 
\usepackage{dsfont}

\def\llama{\texttt{OpenLLaMA-3B}}
\def\pythia{\texttt{Pythia-2.8B}}

\begin{document}
\sloppy
\title{Nob-MIAs: Non-biased Membership Inference Attacks Assessment on Large Language Models with Ex-Post Dataset Construction}

\titlerunning{Nob-MIAs: Non-biased MIAs Assessment on LLMs with Ex-Post Datasets}

\author{C\'edric Eichler\inst{1,2}\textsuperscript{\Letter}\orcidID{0000-0003-3026-1749} \and Nathan Champeil\inst{3,2} \and \\Nicolas Anciaux\inst{2,1}\orcidID{0009-0003-7537-295X} \and Alexandra Bensamoun\inst{4} \and \\ Héber H. Arcolezi\inst{2}\orcidID{0000-0001-8059-7094} \and \\ José Maria De Fuentes\inst{5}\orcidID{0000-0002-4023-3197}}

\institute{LIFO, INSA Centre Val de Loire, Université d'Orléans, Bourges, France\\
\and Inria Saclay, France \email{<firstname.lastname@inria.fr>}\\
\and ENSTA Paris, Palaiseau, France \\
\and Université Paris-Saclay
\email{<alexandra.bensamoun@universite-paris-saclay.fr>}
\and Universidad Carlos III de Madrid, \email{<josemaria.defuentes@uc3m.es>}\\
}
\authorrunning{C. Eichler et al.}
%
%
\maketitle              
\begin{abstract}
The rise of Large Language Models (LLMs) has triggered legal and ethical concerns, especially regarding the unauthorized use of copyrighted materials in their training datasets. This has led to lawsuits against tech companies accused of using protected content without permission. Membership Inference Attacks (MIAs) aim to detect whether specific documents were used in a given LLM pretraining, but their effectiveness is undermined by biases such as time-shifts and n-gram overlaps.

This paper addresses the evaluation of MIAs on LLMs with partially inferable training sets, under the ex-post hypothesis, which acknowledges inherent distributional biases between members and non-members datasets. We propose and validate algorithms to create ``non-biased'' and ``non-classifiable'' datasets for fairer MIA assessment. Experiments using the Gutenberg dataset on OpenLLaMA and Pythia show that neutralizing known biases alone is insufficient. Our methods produce non-biased ex-post datasets on which MIAs achieve AUC-ROC scores comparable to those previously obtained on genuinely random datasets, validating our approach. Globally, MIAs yield results close to random, with only one Meta-Classifier-based MIA being effective on both random and our datasets, but its performance decreases when bias is removed.

\keywords{Membership Inference Attack \and LLM \and Assessment \and Bias.}
\end{abstract}
\section{Introduction}
\label{sec:intro}

The proliferation of Large Language Models (LLMs) has ignited significant legal and ethical debates, particularly concerning copyright infringement. These models often do not document their training data sources, leading to disputes over unauthorized use of copyrighted material. For instance, lawsuits are piling up against OpenAI accused of training ChatGPT using articles and other books without permission\footnote{See, e.g.,  \href{https://www.businessinsider.com/openai-lawsuit-copyrighted-data-train-chatgpt-court-tech-ai-news-2024-6}{``The copyright lawsuits against OpenAI are piling up as the tech company seeks data to train its AI''}, Jun 30, 2024}. Similar accusations have been leveled at Meta or Google for allegedly using protected content\footnote{See, e.g.,  \href{https://www.wired.com/story/congress-senate-tech-companies-pay-ai-training-data/}{``Congress Senate Tech Companies Pay AI Training Data''}, July 2, 2024}. These issues underscore the societal, economic and legal implications of LLM training practices.

Recent surveys highlight the challenges of respecting copyright in AI training across different jurisdictions, such as the USA and France. In the USA, the use of copyrighted data is generally prohibited without the rights holder's permission unless it falls under ``fair use''~\cite{reuel2024open}. For the culture and media sectors, LLM training could not be exempted from this limitation due to the conditions associated with it. More than twenty lawsuits are pending in the USA. In the European Union, the Directive 2019/790 on copyright and related rights in the digital single market introduces a “text and data mining” exception, for any purpose, that could correspond to the use of protected content for LLM training. However, its benefit is conditional on lawful access to copyrighted data and the absence of an opt-out by rights holders. However, not only do the training databases contain infringing works, but most of the rights holders have exercised their opt-out. However, the opacity of the process compromises the return to exclusive rights. So, to provide leverage, the AI Act (European Regulation 2024/1689), the first comprehensive regulation on AI, has required LLM providers, on the one hand, to put in place an internal policy aimed at respecting copyright and, on the other, to be transparent about the sources of training. The AI Office will provide a template on this point. The issue of knowledge of the use of content by the LLM is therefore crucial for rights holders.

\emph{Objective.}
From a technical perspective, determining whether a specific document was a member of the training set of a machine learning (ML) model based on the model's output, is a Membership Inference Attacks (MIAs) problem, highlighted in 2017 by Shokri et al.~\cite{shokri2017membership}. However, the effectiveness of MIAs in the context of LLMs is subject to debate. 

\emph{Limits of existing solutions.}
To determine whether a particular document has been used to train an AI model, MIAs rely on overfitting, which results in stronger predictions when applied to training data. While some research shows that MIAs can achieve high accuracy~\cite{MeeusJRM24-neuron,maini2024llm}, other studies~\cite{Duan2024-doMIAwork,das2024blind} question their validity due to inherent biases in the datasets of members and non-members used for their assessment. For example, biases like time shifts and n-gram overlaps can lead to over-interpretation of results~\cite{Duan2024-doMIAwork}. Additionally, studies indicate that some biases can be exploited to make a ``blind'' classifier, without model access, more effective than MIAs~\cite{das2024blind}. This raises doubts about the robustness and practical relevance of current MIA techniques, and recent surveys like~\cite{JEDRZEJEWSKI2024103988} point out the lack of rigor and practical relevance of current proposals.

\emph{Research question and contributions.} 
This paper aims to address the problem of assessing MIA effectiveness on LLMs which do not disclose their full training datasets but where part of the training dataset can be inferred. The research question we seek to answer is: \textit{``How can we construct an unbiased dataset for evaluating MIAs on LLMs with partially inferable training sets?''} 

We propose and evaluate two approaches: (1) Creating datasets that are ``non-biased'' by design with respect to known biases, and (2) Constructing datasets that cannot be classified, ensuring fairer assessment. Our contributions can be summarized as follows:
\begin{itemize}
    \item We provide algorithms for constructing ex-post datasets of two types: $No-Ngram$ (``No N-gram bias'') and $No-Class$ (``non classifiable''), each designed to mitigate specific types of biases for MIA assessment.
    \item We validate our algorithms and compare our proposed methods in the assessment of existing MIAs.
    \item We demonstrate that neutralizing known biases (e.g., time shifts, n-gram biases) is insufficient for accurate MIA assessment; we also show that several existing MIAs, which are presumed to be effective, are not efficient when evaluated using non-classifiable datasets. For instance, our experiments show that the TPR@10\%FPR and ROC AUC of the best performing MIA out of the 6 assessed drop by 40\% and 14.3\% respectively when evaluated on datasets produced with our approach rather than on randomly sampled ones.
\end{itemize}

\emph{Outline.} 
Section~\ref{sec:rw} reviews related work and positions our research. Section~\ref{sec:pb} synthesizes our hypotheses and the addressed problem. Section~\ref{sec:proposal} presents our proposed solutions and algorithms, detailing ex-post (i.e., a posteriori) construction of unbiased datasets. Section~\ref{sec:XP} provides a comparative experimental evaluation of the proposed solutions. Finally, Section~\ref{sec:conclusion} concludes the paper and suggests directions for future research.

\section{Related Work and Positioning}
\label{sec:rw}
MIA techniques, originally developed for machine learning classification algorithms~\cite{shokri2017membership}, have recently been adapted to the context of LLMs. The baseline technique use \emph{likelihood-based} metrics such as loss~\cite{yeom2018privacy} and perplexity~\cite{carlini2021extracting} to distinguish between \emph{members} (documents which were used for LLM training) and \emph{non-members} (not used in training). 
Several studies show that likelihood-based MIAs applied to LLMs are effective, with high AUC-ROC values. For example,~\cite{MeeusJRM24-neuron} reports an AUC-ROC of $0.856$ for the Gutenberg dataset~\cite{PG19} (also used for evaluation in our paper).
Other metrics, such as Min-k\%Prob~\cite{shi2024detecting}, are based on the premise that if the text has been read by the LLM, it is more likely to appear. 
More precisely, Min-k\%Prob selects the k\% tokens of the document with the minimum probabilities returned by the LLM and computes their average log-likelihood. 

Neighboring-based MIAs calibrate the perplexity score using either neighboring models $\mathcal{M}'$ or neighboring documents $D'$. Classification between members and non-members is then obtained by comparing the likelihood of the target model $\mathcal{M}$ and document $D$ with that of the neighboring model $\mathcal{M}'$ and document $D$, or the model $\mathcal{M}$ and neighboring documents $D'$. Neighboring models hence assume access to a reference model trained on a disjoint data set drawn from a similar distribution, which is often unrealistic~\cite{Duan2024-doMIAwork}. Neighboring documents~\cite{galli2024noisy} are more realistic but present slightly lower performance and additional difficulties in correctly setting noise parameters.

In cases where LLMs do not output likelihood information, complementary metrics can be used with acceptable performance penalty. For example,~\cite{kaneko2024sampling} proposes a MIA method called SaMIA, which measures the similarity between input samples from a document $D$ and the rest of the text in $D$ using ROUGE~\cite{lin2004rouge}. SaMIA demonstrates an AUC-ROC of $0.64$ on subsets of The Pile dataset~\cite{thepile}.

Recent studies~\cite{Duan2024-doMIAwork,maini2024llm} challenge the high-performance claims of MIAs on LLMs. They identify several biases that may skew results, such as timeshift between members and non-members, leading to different distributions of dates and word usage. Re-evaluating some MIAs on Pythia~\cite{pythia}, trained on genuinely random train/test splits of The Pile~\cite{thepile} (and hence have no bias), shows decreased AUC-ROC measures, questioning the apparent success of some MIAs.

Some studies suggest that naive classifiers can distinguish members from non-members with good results based on these biases~\cite{das2024blind,meeus2024inherent}. These studies conclude that MIAs must be evaluated on random datasets taken from a same distribution. While this is possible with open LLMs like Pythia, which reveal their training and test data sources, it is not feasible for LLMs that do not disclose their sources (those of interest in copyright cases).  As shown in the literature (see, e.g.,~\cite{Duan2024-doMIAwork,maini2024llm,meeus2024copyright}), the same MIAs yield different performance (accuracy and relevance) results on different LLMs/datasets. Therefore, the assessment of MIAs on open LLMs cannot be directly transposed to LLMs that do not disclose their training dataset. This confirms the need for techniques like the one we propose. 

A technique inspired by the Regression Discontinuity Design from causal inference, originally used to study treatment effects based on a cutoff date, is proposed in~\cite{meeus2024inherent}. However, documents added just before or after the cutoff date must be known and sufficiently numerous. Additionally, declared dates often deviate from reality~\cite{cheng2024dated}, making this approach impractical.

Many other works are based on MIA attacks on LLMs but rely on different hypotheses, leading to solutions not applicable to our context.~\cite{li2024digger} introduces a framework using loss gap variation during fine-tuning to detect if a document has been seen, though this is not generalizable to initial training documents and also requires an assesment using unbiased datasets. Related works on copyright aspects include copyright issues in LLM outputs~\cite{panaitescu2024can,liu2024shield} to reduce copyrighted text generation and protect users from potential plagiarism, and watermarking techniques~\cite{meeus2024copyright,wei2024proving} for detecting violations in LLM pretraining data, or LLM fine-tuning data~\cite{yan2024protecting}, but these works are not transposable to our context.

\section{Problem Statement}
\label{sec:pb}

A Membership Inference Attack on a large language model $\mathcal{M}$ is a binary classification task aimed at determining whether a specific textual document $D$ was included in the training dataset $\mathcal{D}_{\text{train}}$ used to build $\mathcal{M}$. The goal of this attack is to design a function $MIA: \mathcal{D} \rightarrow \{0,1\}$ that can ascertain the truth value of $D \in \mathcal{D}_{\text{train}}$ for any document in the document space $\mathcal{D}$.

In the context of copyright checks, our goal is to detect ex-post potential violations involving protected texts in the LLM's pretraining dataset. Our hypotheses H1 to H3 stem from this context, acknowledging that the LLM may try to obscure the use of these texts:
\begin{description}
    \item[H1: Self-Assessment.]We assume that a reliable assessment of the MIA must be performed on the target LLM $\mathcal{M}$ itself. As shown in the literature (see, e.g.,~\cite{Duan2024-doMIAwork,maini2024llm}), the same MIAs yield different performance (accuracy and relevance) results on different LLMs/datasets. 
    \item[H2: Partial Member Knowledge.]The training dataset $\mathcal{D}_{\text{train}}$ of $\mathcal{M}$ is partially inferable, i.e., a subset $\mathcal{D}^{known}_{\text{train}}$ of $\mathcal{D}_{\text{train}}$ can be inferred by an attacker. For example, it is know that OpenAI models like GPT-4 have memorized some precise collections of copyright protected books~\cite{chang2023speak}. 
    \item[H3: Bias Recognition.]A subset $\mathcal{D}^{known}_{nm}$ of non-members (i.e., $\mathcal{D}^{known}_{nm} \subset \mathcal{D} \backslash \mathcal{D}_{\text{train}}$) is (obviously) known. Traditionally, such a subset is constructed by considering documents not available at the time the target LLM was released.  
    This creates inherent biases in the ex-post context, where members and non-members are not drawn from the same random distribution. 
\end{description}

We address the problem of producing datasets of members $\mathcal{D}^{Nob}_{train}\subset \mathcal{D}^{known}_{train}$ and non-members $\mathcal{D}^{Nob}_{nm} \subset \mathcal{D}^{known}_{nm}$ which aim to minimize bias (hence the name ``Nob'' for Non-biased). These datasets ensure a reliable evaluation of MIAs on LLMs while satisfying these three hypotheses.

\section{Neutralizing Bias in Ex-Post Dataset Construction}
\label{sec:proposal}

In this section, we present our approach to identifying and mitigating specific bias in the construction of datasets used for the assessment of MIAs. Our methodology operates in two phases: first, addressing bias caused by low n-gram overlap, which has been shown to significantly affect the assessment of MIAs~\cite{Duan2024-doMIAwork}; and second, mitigating additional biases that go beyond n-gram overlap.  

\subsection{Methodology for Identifying and Mitigating Bias}

We begin by targeting n-gram bias, as previous work has demonstrated that n-gram overlaps between members and non-members can distort MIA benchmarks~\cite{Duan2024-doMIAwork}. To counteract this, we propose the $No-Ngram$ algorithm, which aims to generate members and non-members sets with similar distributions of n-gram overlaps.

Next, we leverage traditional classifiers, which we refer to as ``LLM-Agnostic'' classifiers, to identify and mitigate bias beyond n-gram overlap. These classifiers operate without any prior knowledge of the target language model $\mathcal{M}$ or the training dataset $\mathcal{D}_{train}$. 
Our approach uses these classifiers to create member and non-member datasets that resist effective classification. The $No-Class$ algorithm further neutralizes detectable biases, hindering the classifier's ability to distinguish between members and non-members.

\subsection{Neutralizing N-gram Bias}
\label{sec:n-gramProposal}

The impact of n-gram distribution on MIA performance has been extensively documented. For example, time-shifted datasets often exhibit variations in n-gram distribution due to changes in dates, but also language, vocabulary and topics of interest over time~\cite{Duan2024-doMIAwork}. A significant difference in n-gram overlap between non-members and left-out members can lead to an inflated evaluation of MIA performance. To mitigate this, we propose the $No-Ngram$ algorithm (see Algorithm~\ref{alg:nongram}), which generates member and non-member sets with distributions of n-gram overlap w.r.t. left-out members that closely match.

\begin{algorithm}[!]
\caption{$No-Ngram$}\label{alg:nongram}
\textbf{Input:} $\mathcal{D}^{known}_{train}$ set of known members, $\mathcal{D}^{known}_{nm}$ set of non-members, integer $n$ the size of the output datasets, $Dist$ distance between two vectors \\
\textbf{Output:} $\mathcal{D}^{Nob}_{train} \subset \mathcal{D}^{known}_{train}$ a set of members, $\mathcal{D}^{Nob}_{nm}\subset \mathcal{D}^{known}_{nm}$ non-members, minimizing N-gram bias
\begin{algorithmic}[1]
\Require $n \leq 1/2*|D^{known}_{train}|$

\Ensure $n = |\mathcal{D}^{Nob}_{train}| =  |\mathcal{D}^{Nob}_{nm}|$

\State $\mathcal{D}^{Nob}_{train} \gets $ random sample of $\mathcal{D}^{known}_{train}$ of size $n$ 

\State $\mathcal{D}_{train}^{remain} \gets \mathcal{D}^{known}_{train} \backslash \mathcal{D}^{Nob}_{train} $
\State $distrib_{train} \gets distribution(\mathcal{D}^{Nob}_{train},\mathcal{D}^{remain}_{train})$ \Comment{Compute n-gram overlap distribution}
\State $\mathcal{D}^{Nob}_{nm}\gets \emptyset$
\For{$i=1$ to $n$} \Comment{Select document minimizing overlap distributions distance}
\State $\mathcal{D}_{nm}^{remain} \gets \mathcal{D}^{known}_{nm} \backslash \mathcal{D}^{Nob}_{nm} $
\State $D \gets \underset{D \in \mathcal{D}_{nm}^{remain}}{\arg\min}(d_{Kolmogorov-Sminrov}(distribution(\{D\} \cup \mathcal{D}^{Nob}_{nm}), distrib_{train})$
\State $\mathcal{D}^{Nob}_{nm}\gets \{D\} \cup \mathcal{D}^{Nob}_{nm}$
\EndFor
\end{algorithmic}
\end{algorithm}

\noindent\textbf{Algorithm overview.} Algorithm~\ref{alg:nongram} operates in three steps:
\begin{itemize}
    \item \textit{Initial sampling:} The algorithm begins by selecting an arbitrary sample $\mathcal{D}^{Nob}_{train}$ of $\mathcal{D}^{known}_{train}$ of an appropriate size (line 1). 
    \item \textit{Overlap distribution computation:} It then computes the distribution of n-gram overlap between the selected members and remaining ones (line 3) using the function $distribution$ (described below). This distribution represents the target overlap distribution that the non-member set should mirror to be indistinguishable from selected members. 
    \item \textit{Greedy construction:} Afterwards, the non-member dataset is constructed document by document, in a greedy fashion: at each step (lines 5 to 9), the document that minimizes the Kolmogorov-Smirnov distance (see below) between the n-gram overlap distributions is selected. 
\end{itemize}

The Kolmogorov-Smirnov (KS) distance\footnote{Other distance metrics could be used. Exploring them is planned for future work.} used in the algorithm is a widely used metric for measuring the distance between (real, non parametric) distributions. In the context of LLMs, it is particularly useful for comparing distributions of generated verbatim text as it appears in the training data or prompts~\cite{sonkar2024many}. Other distance could be used, which is considered future work. 

The $distribution$ function produces the distribution of n-gram overlap of a dataset with reference to another. For a given document $D$, which is considered as a sequence of $k$ tokens (such as letters or words), an n-gram is defined as a continuous sequence of $n$ tokens. The overlap of n-grams from a document $D$ with reference to a set of documents $\mathcal{D}_{ref}$ is computed as the percentage of n-grams in $D$ that appear in any document of $\mathcal{D}_{ref}$. The resulting distribution is comprised of the overlap scores of each document in the first dataset.

\subsection{Constructing a Non-Classifiable Dataset}

To mitigate bias indicated by the ability of agnostic classifiers to distinguish between members and non-members, we introduce the $No-Class$ algorithm (see Algorithm~\ref{alg:noClass}). This algorithm is designed to produce datasets where the performance of classifiers is minimized, effectively neutralizing their ability to differentiate between members and non-members.

\begin{algorithm}
\caption{$No-Class$ Dataset Generation}\label{alg:noClass}
\textbf{Input} $\mathcal{D}^{known}_{train}$, $\mathcal{D}^{known}_{nm}$, integer $n$ the size of the output datasets, $(C_i)_{i \in [1,N]}$ a vector of agnostic classifiers outputing $\mathds{P}[C_i(D)]$ the probability of $D$ being a member \\
\textbf{Output} $\mathcal{D}^{Nob}_{train} \subset \mathcal{D}^{known}_{train}$, $\mathcal{D}^{Nob}_{nm}\subset \mathcal{D}^{known}_{nm}$
\begin{algorithmic}[1]
\Require $n \leq 1/4 \times|D^{known}_{train}| \And n \leq 1/4\times|D^{known}_{nm}|$
\State $\mathcal{D}_{m} \gets $ random sample of $\mathcal{D}^{known}_{train}$ of size $n$ 
\State $\mathcal{D}_{nm} \gets $ random sample of $\mathcal{D}^{known}_{nm}$ of size $n$ 
\State train each $(C_i)_{i \in [1,N]}$ on $\mathcal{D}_{m} \cup \mathcal{D}_{nm}$

\State $\mathcal{D}_{train}^{Nob} \gets \emptyset$
\State $\mathcal{D}^{remain}_{train}\gets \mathcal{D}^{known}_{train} \backslash  \mathcal{D}_{m}$
\State $\mathcal{D}_{nm}^{Nob} \gets \emptyset$
\State $\mathcal{D}^{remain}_{nm}\gets \mathcal{D}^{known}_{nm} \backslash  \mathcal{D}_{nm}$
\For{$i=1$ to $n$} \Comment{populating members and non-members minimizing confidence}
\State $D_m \gets \underset{D_m \in \mathcal{D}_{remain}^{train}}{\arg\min} \left \lVert (C_i(D_m)-0.5)_{i \in [1,N]} \right \rVert_2$
\State $D_{nm} \gets \underset{D_{nm} \in \mathcal{D}_{remain}^{nm}}{\arg\min} \left \lVert (C_i(D_{nm})-0.5)_{i \in [1,N]} \right \rVert_2$
\State $\mathcal{D}_{train}^{Nob} \gets \mathcal{D}_{train}^{Nob} \cup \{D_m\}$
\State $\mathcal{D}^{remain}_{train}\gets \mathcal{D}^{remain}_{train} \backslash \{D_m\}$
\State $\mathcal{D}_{nm}^{Nob} \gets \mathcal{D}_{nm}^{Nob} \cup \{D_{nm}\}$
\State $\mathcal{D}^{remain}_{nm}\gets \mathcal{D}^{remain}_{nm} \backslash \{D_{nm}\}$
\EndFor
\end{algorithmic}
\end{algorithm}

\noindent\textbf{Algorithm overview.} In Algorithm~\ref{alg:noClass}, we consider a vector of $N$ classifiers $(C_i)_{i \in [1,N]}$, each of which, once trained, assigns a probability in the range $[0,1]$ to indicate the likelihood of a document being a member. The closer to $1$ (respectively, $0$), the more confident the classifier is that the document is a member (resp., non-member). The intuition behind our algorithm is to exploit the confidence to ensures that the constructed datasets are as challenging as possible for the classifiers. It is worth noting that other variants of this algorithm have been implemented, balancing the number of false positives, false negatives, true positives, and true negatives in each member/non-member class. The algorithm operates in two main steps: 
\begin{itemize} 

    \item \textit{Sampling and training:} The algorithm begins by randomly sampling known members and non-members from the dataset (lines 1-2). These samples are then used to train a set of $N$ agnostic classifiers $(C_i)_{i \in [1,N]}$ (line 3).
    
    \item \textit{Confidence minimization:} Using the classifiers $(C_i)_{i \in [1,N]}$ on the left-out members, we then construct $\mathcal{D}_{train}^{Nob}$ and $\mathcal{D}_{nm}^{Nob}$ (lines 8 to 14) minimizing the overall confidence of the classifiers. Since the further $C_i(D)$ is from $0.5$, the more confident $C_i$ is in its assessment of document $D$, we consider the vector $(C_i(D)-0.5)_{i \in [1,N]}$ representing the confidence of each classifiers. At each step, we add the element $D$ that minimizes the l2-norm of this confidence vector. For instance, considering the construction of considered members: (1) among the members that have neither been selected as ``Non-biased'' (Nob) nor used to train the classifiers ($D \in \mathcal{D}^{remain}_{train}$), the one that minimizes the l2-norm of the confidence vector (i.e., $\left \lVert (C_i(x)-0.5)_{i \in [1,N]} \right \rVert_2$) is selected (line 9); (2) the aforementioned element is inserted in the set (line 11); (3) the set of remaining candidates is updated (line 12).
\end{itemize}

\section{Experimental Validation of Our Approach}
\label{sec:XP}

In this section, we apply and evaluate our proposal with reference to the Gutenberg dataset. Experimental settings are described in Sec.~\ref{sec:settings}, detailing how the candidate datasets are constructed to avoid a priori bias, how bias are assessed, as well as the MIAs and LLMs assessed on the datasets produced following our proposal. In spite of known members and non-members being constructed to circumvent bias, Sec.~\ref{sec:rand} shows that random samples exhibit n-gram bias. Such bias are addressed in Sec~\ref{sec:xpGram} by producing datasets following the $No-Ngram$ algorithm (Alg.~\ref{alg:nongram}), which still exhibit residual bias exploitable by an agnostic classifier. Section~\ref{sec:xpNoClass} assesses the last pair of datasets produced following the $No-Class$ algorithm (Alg.~\ref{alg:noClass}). Finally, Sec.~\ref{xp:AssessMIA} presents the assessment of MIAs using the produced datasets and discusses the impact of bias in the evaluation of MIAs. All the code, resulting analysis and dataset are available online\footnote{\url{https://github.com/ceichler/MIA-bias-removal}}.

\subsection{Experimental Setting}
\label{sec:settings}

\noindent\textbf{Dataset: Gutenberg Project.}
The project Gutenberg\footnote{\url{https://www.gutenberg.org/}} offers a high-quality open dataset of over 70,000 books, continuously expanding. PG-19~\cite{PG19}, a subset of 28,752 books extracted in 2019, has been included in RedPajama-Data~\cite{redpajama} and The Pile~\cite{thepile} and used to train LLMs such as Pythia~\cite{pythia} and OpenLLaMA~\cite{openllama}. It is also widely used to evaluate MIAs (e.g.,~\cite{MeeusJRM24-neuron},~\cite{maini2024llm}). We use documents from Project Gutenberg in our experiments because of its quality, recognized relevance in MIA research and the availability of methods~\cite{MeeusJRM24-neuron} to minimize bias.

We assume an LLM trained on PG-19~\cite{PG19} and draw our \emph{members} from this dataset. Regarding \emph{non-members}, since Project Gutenberg is ongoing, with books continuously added, all English books added after the publication of PG-19 are potential non-members. To circumvent the potential for temporal bias between the member and non-member sample, we adhere to the methodology proposed by Meeus et al.~\cite{MeeusJRM24-neuron} and restrict our analysis to books published between 1850 and 1910. This leads to final sets $\mathcal{D}^{known}_{\text{train}}$ and $\mathcal{D}^{known}_{nm}$ of 7300 and 2400 books, respectively. Note also that Das et al.~\cite{das2024blind} identified a potential bias in datasets constructed following this methodology. Indeed, they showed that the format of the preface metadata that project Gutenberg adds to books has changed since 2019. To circumvent this, we discard such metadata.  Therefore, our starting sets of members $\mathcal{D}^{known}_{\text{train}}$ and non-members $\mathcal{D}^{known}_{nm}$ are chosen because, \textbf{a priori, there is no (known) bias affecting them.}
~\\

\noindent\textbf{Bias assessment.} The \emph{agnostic classifiers} employed in this study utilize a Bayes algorithm for multinomially distributed data, focusing on the distribution of 1 to 3-grams. These classifiers are trained and applied using scikit-learn, chosen for its robustness in handling such data distributions.
For the \emph{n-gram analysis}, characters are treated as tokens when computing n-grams. 
We focus on n=7, as previous research~\cite{Duan2024-doMIAwork} has shown that 7-grams reveal the most significant distributional differences.
The n-gram analysis is conducted using a Bloom filter based on the implementation of~\cite{bff}, ensuring efficient and accurate bias detection. 
~\\

\noindent\textbf{LLMs.} We conduct our experiments on two autoregressive large language models: OpenLLaMA~\cite{openllama} and Pythia~\cite{pythia}. 
OpenLLaMA is a series of 3B, 7B and 13B open-source models trained on 1T tokens that aims to emulate Meta's LLaMA~\cite{llama}. OpenLLaMA is trained on RedPajama-Data~\cite{redpajama}, an open-source reproduction of the original LLaMA training dataset. 
Pythia is an open and transparent suite of LLMs ranging in size from 70M to 12B parameters that has been specifically released to enable research. The language models in Pythia have been trained on The Pile~\cite{thepile}.
In this work, we have used the \llama{} and \pythia{} models. 
Both The Pile and RedPajama-Data include PG-19~\cite{PG19}.
~\\

\noindent\textbf{MIAs.} We conduct our experiments adapting the codes provided by~\cite{maini2024llm} with the following state-of-the-art MIAs:

\begin{itemize}
    \item \textbf{Min-k\% Prob} is based on the likelihood of the \( k\% \) of tokens in a sequence \( D \) that have the lowest probabilities, based on the preceding tokens~\cite{shi2024detecting}. 

    \item \textbf{Max-k\% Prob} is the inverse metric of Min-k\% Prob, based on the tokens that have the highest probabilities. We use \( k = 10 \) for both Min-k\% Prob and Max-k\% Prob.

    \item \textbf{zlib Ratio} identifies potential member when having a low ratio of the model's perplexity to the entropy of the text~\cite{carlini2021extracting}. This entropy is calculated as the number of bits required to compress the sequence using~\cite{gailly2004zlib}.

    \item \textbf{Perplexity (ppl)} leverages perplexity~\cite{carlini2021extracting} as scores and then threshold them to classify samples as members or non-members.
    
    \item \textbf{Meta\_MIA} is based on the work of~\cite{maini2024llm}, which aggregates 52 MIAs (including Min-k\% Prob, perplexity, zlib Ratio, etc.) to create a single feature vector.
    A linear regressor is trained to learn the importance of weights for the different MIA attacks and thus classify their membership status.

\end{itemize}
\subsection{Assuming No bias: Random Sample}
\label{sec:rand}
\begin{figure*}[h!]
    \centering
    \begin{subfigure}[t]{0.5\textwidth}
        \centering
        \includegraphics[clip, trim=0cm 0.2cm 0cm 1.4cm,width=\textwidth]{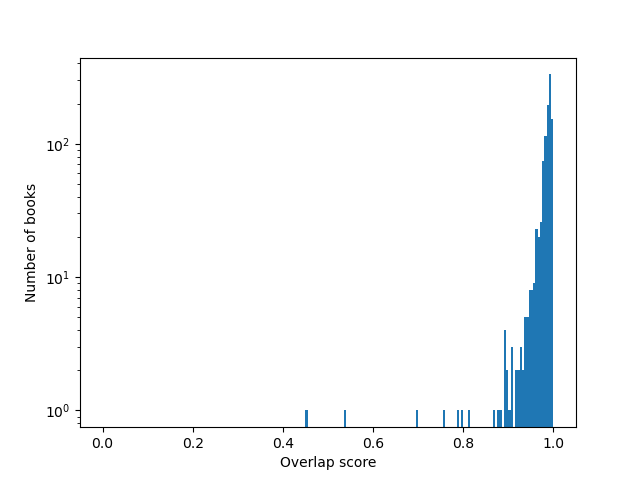}
        \label{fig:m1mngram}
        \caption{Randomly selected members.}
    \end{subfigure}%
    ~ 
    \begin{subfigure}[t]{0.5\textwidth}
        \centering
    \includegraphics[clip, trim=0cm 0.2cm 0cm 1.4cm,width=\textwidth]{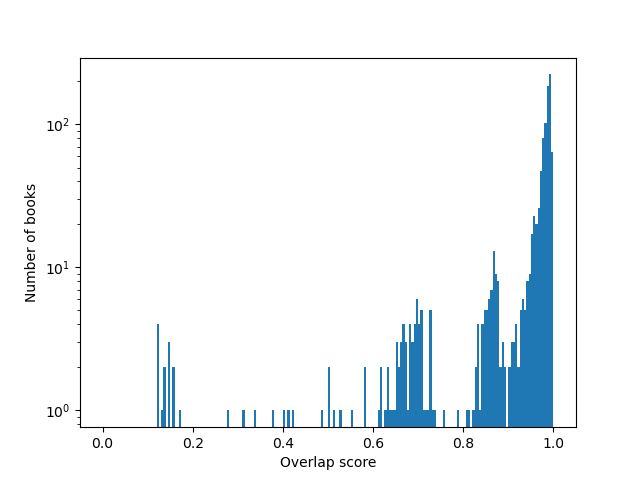}
        \label{fig:m2nmngram}
        \caption{Randomly selected non-members.}
    \end{subfigure}
    \caption{Histograms of n-gram overlap w.r.t. left out members (KS-dist = 0.222).}
    \label{fig:ngramHistM1}
\end{figure*}

By construction, $\mathcal{D}^{known}_{train}$ and $\mathcal{D}^{known}_{nm}$ are exempt of bias related to meta-data and time-shift. Since there is no reason to suspect a bias, we construct $\mathcal{D}^{Nob}_{train}$ and $\mathcal{D}^{Nob}_{nm}$ through a random sample. We compute the distributions of n-gram overlap of these two sets with reference to the left out members ($\mathcal{D}^{known}_{train} \backslash \mathcal{D}^{Nob}_{train}$) as described in Sec.~\ref{sec:n-gramProposal}. The result is depicted as histograms in Fig.~\ref{fig:ngramHistM1}.

Surprisingly, in-spite of the absence of time-shift and metadata bias, the set of members and non-members exhibit significant distributional difference of n-gram overlap, with a KS distance of 0.222. This bias can be exploited by an agnostic classifier achieving an AUC ROC of 0.84 (reported hereafter).
 
 \noindent\textit{Conclusion.}  The text of books written in the same time interval but added to project Gutenberg at different dates (before or after the extraction of PG-19) still exhibit n-gram shifts and a random sampling produce heavily biased datasets.

\subsection{$No-Ngram$ to Minimize N-gram Bias}
\label{sec:xpGram}
\begin{figure*}[t!]
    \centering
    \begin{subfigure}[t]{0.5\textwidth}
        \centering
        \includegraphics[clip, trim=0cm 0cm 0cm 1.4cm,width=\textwidth]{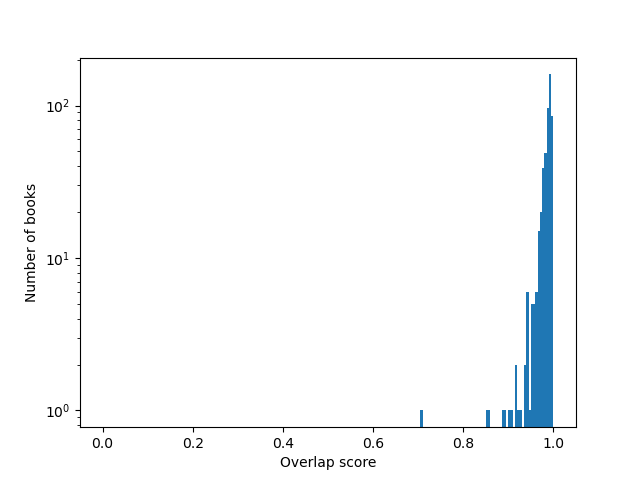}
        \caption{$No-Ngram$ members.}
        \label{fig:m2mngram}
    \end{subfigure}%
    ~ 
    \begin{subfigure}[t]{0.5\textwidth}
        \centering
    \includegraphics[clip, trim=0cm 0cm 0cm 1.4cm,width=\textwidth]{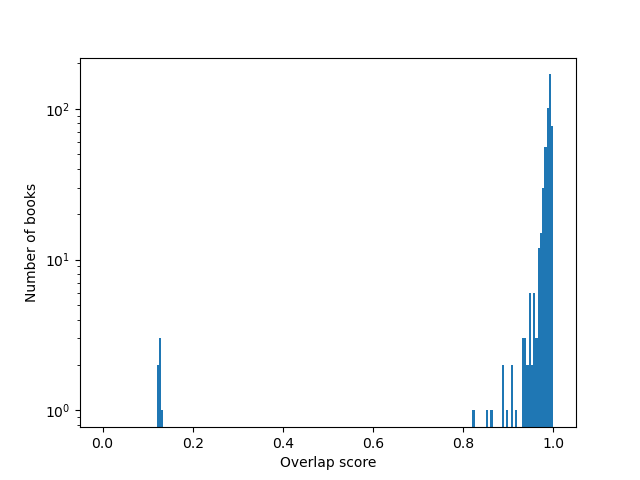}
        \caption{$No-Ngram$ non-members.}
        \label{fig:m2nmngram}
    \end{subfigure}
    \caption{Histograms of n-gram overlap w.r.t. left out members (KS-dist = 0.034).}
    \label{fig:m2ngram}
    \centering
    \begin{subfigure}[t]{0.5\textwidth}
        \centering
        \includegraphics[clip, trim=0cm 0cm 0cm 1.4cm, width=\textwidth]{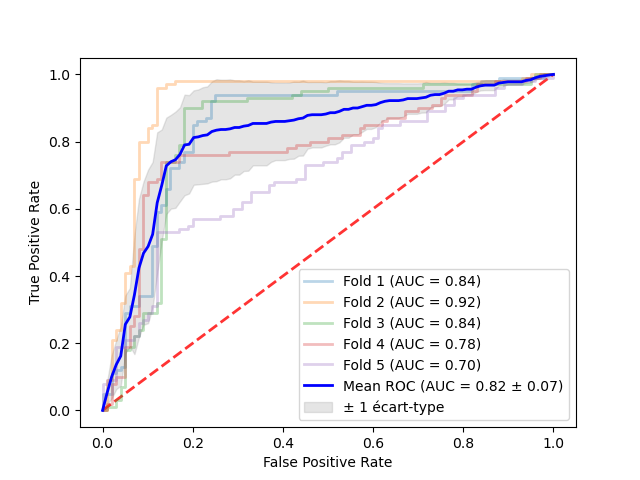}
        \caption{Trained on $No-Ngram$ sets.}
        \label{fig:rocM2}
    \end{subfigure}%
    ~ 
    \begin{subfigure}[t]{0.5\textwidth}
        \centering
    \includegraphics[clip, trim=0cm 0cm 0cm 1.4cm,width=\textwidth]{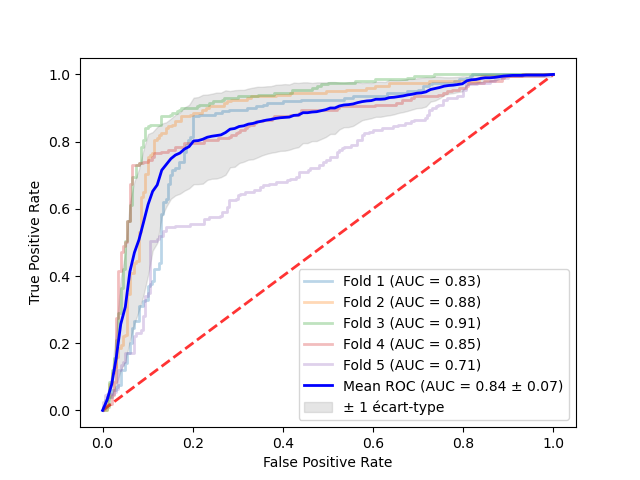}
        \caption{Trained on randomly sampled set.}
        \label{fig:rocM1}
    \end{subfigure}
    \caption{ROC of 5-folds agnostic classifiers.}
\end{figure*}

To address the highlighted n-gram overlap bias, we produce new samples $\mathcal{D}^{Nob}_{train}$ and $\mathcal{D}^{Nob}_{nm}$ following the $No-Ngram$ algorithm. The corresponding distributions of n-gram overlap are depicted in Fig.~\ref{fig:m2ngram}. Their KS distance is 0.034, a drastic 84\% drop from 0.222, the distance achieved with random samples. Notably, non-members exhibit high n-gram overlap with the left out members, only 5 non-member books having an overlap score lesser than 0.8.

We further assess residual bias by training a classifier on $\mathcal{D}^{Nob}_{train}$ and $\mathcal{D}^{Nob}_{nm}$. The ROC of each fold is illustrated in Fig.~\ref{fig:rocM2}. As a reference, the evaluation of a classifier trained on randomly sampled datasets is depicted in Fig.~\ref{fig:rocM1}. The agnostic classifier achieves on average over 5 folds 0.82 AUC ROC and 5\%, 26\%, and 49\%TPR at 1\%, 5\%, and 10\% FPR, respectively. This remains highly accurate and the accuracy loss is marginal when compared to random samples where an agnostic classifier achieves 0.84 AUC ROC and 3\%, 30\%, and 63\% TPR at 1\%, 5\%, and 10\% FPR, respectively.

\noindent\textit{Conclusion.}  While the $No-Ngram$ algorithm has successfully reduced (if not eliminated) the bias in n-gram overlap, the produced sets can still be discriminated with high accuracy by an agnostic classifier. Contrarily to previous proposal~\cite{Duan2024-doMIAwork}, this suggests that distributional difference in n-gram overlap is insufficient as a metric of MIA benchmark difficulty.
\subsection{Sampling Unclassifiable Datasets}
\label{sec:xpNoClass}
\begin{figure*}[h!]
        \centering
        \includegraphics[clip, trim=0cm 0cm 0cm 1.4cm, width=0.5\textwidth]
        {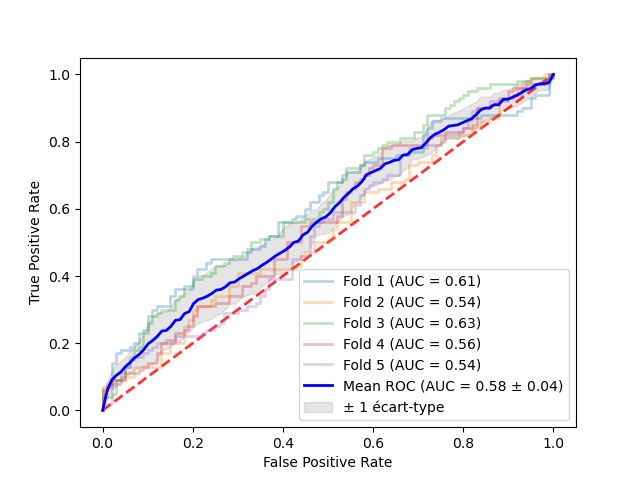}
        \caption{ROC of 5-folds agnostic classifier trained on $No-Class$ sets.}
        \label{fig:roc3}
\end{figure*}

The datasets being classifiable even when n-gram bias are minimized, we apply $No-Class$\footnote{Since we use a single classifier, we also ensure the same number of false positive, false negative, true positive and true negative in the selected sets.} using a single classifier trained on randomly sampled set whose evaluation is presented in Fig.~\ref{fig:rocM1}. To evaluate residual bias, we train a new agnostic classifier on the resulting sets whose evaluation is shown in Fig.~\ref{fig:roc3}.

On average, the agnostic classifier achieves 6\%, 13\%, and 20\%TPR at 1\%, 5\%, and 10\% FPR, respectively. Interestingly, the 5th fold achieves lower TPR than a random guess for FPR in the 30-45\% interval. 
Overall, performance is slightly better than random, particularly at low FPR, but significantly worse than previous settings.
Indeed, the TPR at 5\% and 10\% FPR decrease by roughly 56\% and 72\% when compared to a classifier trained on random samples, respectively. Similarly, the AUC ROC drops from 0.84 to 0.58, denoting a 76\% decrease of the distance to the AUC ROC of a random guess. 

\noindent\textit{Conclusion.} These results indicate that hard-to-classify sets also resist training, showing minimal exploitable bias for agnostic classifiers.

\subsection{Assesment of MIAs}
\label{xp:AssessMIA}

We assess 6 state of the art MIAs on 2 LLMs and the datasets random, $No-Ngram$, and $No-Class$ presented in Sec.~\ref{sec:rand},~\ref{sec:xpGram}, and~\ref{sec:xpNoClass}, respectively. Tables~\ref{tab:tpr_10_fpr} and~\ref{tab:roc_auc_results} report their TPR at 10\%FPR and AUC ROC averaged over 5 runs.

\begin{table}[tb]
    \centering
    \begin{tabular}{ccccccc}
    \hline
    Model & Dataset & Meta\_MIA & Max10\%Prob & ppl & zlib\_ratio & Max10\%Prob \\
    \hline \hline
    \pythia & random & 0.314 & 0.085 & 0.087 & 0.111 & 0.062 \\
    \pythia & $No-Ngram$ & 0.350 & 0.149 & 0.050 & 0.145 & 0.124 \\
    \pythia & $No-Class$ & \textbf{0.314} & \textbf{0.313} & 0.127 & 0.140 & 0.162 \\
    \hline
    \llama & random & 0.371 & 0.111 & 0.059 & 0.139 & 0.122 \\
    \llama & $No-Ngram$ & 0.386 & 0.156 & 0.074 & 0.131 & 0.222 \\
    \llama & $No-Class$ & \textbf{0.224} & 0.070 & 0.046 & 0.132 & 0.081 \\
    \hline
    \end{tabular}
    \caption{TPR values at 10\%FPR. Bold values outperform agnostic classifiers.}
    \label{tab:tpr_10_fpr}

    \centering
    \begin{tabular}{ccccccc}
    \hline 
    Model & Dataset & Meta\_MIA & Min10\%Prob & ppl & zlib\_ratio & Max10\%Prob \\
    \hline \hline
    \pythia & random & 0.692 & 0.544 & 0.554 & 0.494 & 0.490 \\
    \pythia & $No-Ngram$ & 0.688 & 0.538 & 0.477 & 0.544 & 0.557 \\
    \pythia & $No-Class$ & 0.670 & 0.665 & 0.583 & 0.531 & 0.578 \\
    \hline
    \llama & random & 0.740 & 0.523 & 0.493 & 0.501 & 0.576 \\
    \llama & $No-Ngram$ & 0.744 & 0.543 & 0.545 & 0.506 & 0.642 \\
    \llama & $No-Class$ & \textbf{0.634} & 0.520 & 0.503 & 0.523 & 0.494 \\
    \hline
    \end{tabular}

    \caption{AUC ROC values. Bold values outperform an agnostic classifier.}
    \label{tab:roc_auc_results}
\end{table}
Overall, no MIA manage to outperform an agnostic classifier on the random and $No-Ngram$ datasets. Only Meta-MIA outperforms the classifier on $No-Class$ on both \llama{} and \pythia, while 10\%\_min\_probs outperforms it solely on \pythia{} according to the TPR@10\%FPR. 

Meta-MIA is consistently significantly above a random guess and the best MIA across all settings and metric. It achieves its best results on \pythia, with 37.1\% TPR@10\%FPR and a AUC ROC of 0.74 for the random biased dataset. On $No-Class$, these values drop to 22.4\% and 0.634, denoting a ratio $No-Class$/random of 0.60 and 0.857.

\noindent\textit{Conclusion.} Meta-MIA is consistently the best out of the 6 MIAs evaluated on our datasets. Yet, its TPR@10\%FPR and AUC ROC drop by 40\% and 14.3\% respectively when evaluated on datasets produced using our approach rather than on randomly sampled ones. This underlines the importance of our approach to accurately estimate MIAs performances.

\section{Conclusion}
\label{sec:conclusion}
As LLMs are trained leveraging myriads of data items, including copyrighted ones, it is key to ascertain whether a piece of data has been used in this process. Yet, the effectiveness of MIAs has been recently questioned, due to the existence of biases in datasets constructed ex-post. This work introduces $Nob-MIAs$, a set of algorithms to build unbiased datasets, thus setting a more solid ground for MIA assessment. Our experiments on the Gutenberg dataset confirms that our approach significantly reduces bias (e.g., an 84\% reduction of difference in n-gram overlap distribution) and impacts on MIA evaluation, with TPR@10\%FPR and ROC AUC of the best-performing MIA (out of 6) decreasing by 40\% and 14.3\% respectively, compared to evaluations on randomly sampled datasets.

This work opens several future research avenues, including extending the algorithms to detect and mitigate residual biases and applying this approach to non-textual MIAs, where ex-post dataset construction is also common.

\begin{credits}
\subsubsection{\ackname} 

This work was supported by the ``ANR 22-PECY-0002'' \href{https://www.pepr-cybersecurite.fr/projet/ipop/}{IPoP} (Interdisciplinary Project on Privacy) project of the Cybersecurity PEPR and DATAIA. 
Jose Maria de Fuentes has also received support from the Spanish National Cybersecurity Institute (INCIBE) grant APAMciber within the framework of the Recovery, Transformation and Resilience Plan funds, financed by the European Union (Next Generation); and from UC3M's Requalification programme, funded by the Spanish Ministerio de Ciencia, Innovacion y Universidades with EU recovery funds (Convocatoria de la Universidad Carlos III de Madrid de Ayudas para la recualificación del sistema universitario español para 2021-2023, de 1 de julio de 2021).

\end{credits}

\bibliographystyle{splncs04}
\bibliography{ref}

\begin{thebibliography}{10}
\providecommand{\url}[1]{\texttt{#1}}
\providecommand{\urlprefix}{URL }
\providecommand{\doi}[1]{https://doi.org/#1}

\bibitem{pythia}
Biderman, S., Schoelkopf, H., Anthony, Q., Bradley, H., O'Brien, K., Hallahan,
  E., Khan, M.A., Purohit, S., Prashanth, U.S., Raff, E., Skowron, A.,
  Sutawika, L., Van Der~Wal, O.: Pythia: a suite for analyzing large language
  models across training and scaling. In: Proceedings of the 40th International
  Conference on Machine Learning. ICML'23, JMLR.org (2023)

\bibitem{carlini2021extracting}
Carlini, N., Tramer, F., Wallace, E., Jagielski, M., Herbert-Voss, A., Lee, K.,
  Roberts, A., Brown, T., Song, D., Erlingsson, U., et~al.: Extracting training
  data from large language models. In: 30th USENIX Security Symposium (USENIX
  Security 21). pp. 2633--2650 (2021)

\bibitem{chang2023speak}
Chang, K.K., Cramer, M., Soni, S., Bamman, D.: Speak, memory: An archaeology of
  books known to chatgpt/gpt-4. arXiv preprint arXiv:2305.00118  (2023)

\bibitem{cheng2024dated}
Cheng, J., Marone, M., Weller, O., Lawrie, D., Khashabi, D., Van~Durme, B.:
  Dated data: Tracing knowledge cutoffs in large language models. arXiv
  preprint arXiv:2403.12958  (2024)

\bibitem{redpajama}
Computer, T.: Redpajama-data: An open source recipe to reproduce llama training
  dataset (2023), \url{https://github.com/togethercomputer/RedPajama-Data}

\bibitem{das2024blind}
Das, D., Zhang, J., Tram{\`e}r, F.: Blind baselines beat membership inference
  attacks for foundation models. arXiv preprint arXiv:2406.16201  (2024)

\bibitem{Duan2024-doMIAwork}
Duan, M., Suri, A., Mireshghallah, N., Min, S., Shi, W., Zettlemoyer, L.,
  Tsvetkov, Y., Choi, Y., Evans, D., Hajishirzi, H.: Do membership inference
  attacks work on large language models? arXiv preprint arXiv:2402.07841
  (2024)

\bibitem{gailly2004zlib}
Gailly, J.l., Adler, M.: Zlib compression library  (2004)

\bibitem{galli2024noisy}
Galli, F., Melis, L., Cucinotta, T.: Noisy neighbors: Efficient membership
  inference attacks against llms. arXiv preprint arXiv:2406.16565  (2024)

\bibitem{thepile}
Gao, L., Biderman, S., Black, S., Golding, L., Hoppe, T., Foster, C., Phang,
  J., He, H., Thite, A., Nabeshima, N., Presser, S., Leahy, C.: The pile: An
  800gb dataset of diverse text for language modeling (2020)

\bibitem{openllama}
Geng, X., Liu, H.: Openllama: An open reproduction of llama (May 2023),
  \url{https://github.com/openlm-research/open_llama}

\bibitem{bff}
Groeneveld, D., Ha, C., Magnusson, I.: Bff: The big friendly filter (2023),
  \url{https://github.com/allenai/bff}

\bibitem{JEDRZEJEWSKI2024103988}
Jedrzejewski, F.V., Thode, L., Fischbach, J., Gorschek, T., Mendez, D.,
  Lavesson, N.: Adversarial machine learning in industry: A systematic
  literature review. Computers \& Security p. 103988 (2024)

\bibitem{kaneko2024sampling}
Kaneko, M., Ma, Y., Wata, Y., Okazaki, N.: Sampling-based pseudo-likelihood for
  membership inference attacks. arXiv preprint arXiv:2404.11262  (2024)

\bibitem{li2024digger}
Li, H., Deng, G., Liu, Y., Wang, K., Li, Y., Zhang, T., Liu, Y., Xu, G., Xu,
  G., Wang, H.: Digger: Detecting copyright content mis-usage in large language
  model training. arXiv preprint arXiv:2401.00676  (2024)

\bibitem{lin2004rouge}
Lin, C.Y.: Rouge: A package for automatic evaluation of summaries. In: Text
  summarization branches out. pp. 74--81 (2004)

\bibitem{liu2024shield}
Liu, X., Sun, T., Xu, T., Wu, F., Wang, C., Wang, X., Gao, J.: Shield:
  Evaluation and defense strategies for copyright compliance in llm text
  generation. arXiv preprint arXiv:2406.12975  (2024)

\bibitem{maini2024llm}
Maini, P., Jia, H., Papernot, N., Dziedzic, A.: Llm dataset inference: Did you
  train on my dataset? arXiv preprint arXiv:2406.06443  (2024)

\bibitem{MeeusJRM24-neuron}
Meeus, M., Jain, S., Rei, M., de~Montjoye, Y.: Did the neurons read your book?
  document-level membership inference for large language models. In:
  Balzarotti, D., Xu, W. (eds.) 33rd {USENIX} Security Symposium, {USENIX}
  Security 2024, Philadelphia, PA, USA, August 14-16, 2024. {USENIX}
  Association (2024)

\bibitem{meeus2024inherent}
Meeus, M., Jain, S., Rei, M., de~Montjoye, Y.A.: Inherent challenges of
  post-hoc membership inference for large language models. arXiv preprint
  arXiv:2406.17975  (2024)

\bibitem{meeus2024copyright}
Meeus, M., Shilov, I., Faysse, M., de~Montjoye, Y.A.: Copyright traps for large
  language models. In: 41st International Conference on Machine Learning (2024)

\bibitem{panaitescu2024can}
Panaitescu-Liess, M.A., Che, Z., An, B., Xu, Y., Pathmanathan, P., Chakraborty,
  S., Zhu, S., Goldstein, T., Huang, F.: Can watermarking large language models
  prevent copyrighted text generation and hide training data? arXiv preprint
  arXiv:2407.17417  (2024)

\bibitem{PG19}
Rae, J.W., Potapenko, A., Jayakumar, S.M., Lillicrap, T.P.: Compressive
  transformers for long-range sequence modelling. arXiv preprint
  arXiv:1911.05507  (2019)

\bibitem{reuel2024open}
Reuel, A., Bucknall, B., Casper, S., Fist, T., Soder, L., Aarne, O., Hammond,
  L., Ibrahim, L., Chan, A., Wills, P., et~al.: Open problems in technical ai
  governance. arXiv preprint arXiv:2407.14981  (2024)

\bibitem{shi2024detecting}
Shi, W., Ajith, A., Xia, M., Huang, Y., Liu, D., Blevins, T., Chen, D.,
  Zettlemoyer, L.: Detecting pretraining data from large language models. In:
  The Twelfth International Conference on Learning Representations (2024)

\bibitem{shokri2017membership}
Shokri, R., Stronati, M., Song, C., Shmatikov, V.: Membership inference attacks
  against machine learning models. In: 2017 IEEE symposium on security and
  privacy (SP). pp. 3--18. IEEE (2017)

\bibitem{sonkar2024many}
Sonkar, S., Baraniuk, R.G.: Many-shot regurgitation (msr) prompting. arXiv
  preprint arXiv:2405.08134  (2024)

\bibitem{llama}
Touvron, H., Lavril, T., Izacard, G., Martinet, X., Lachaux, M.A., Lacroix, T.,
  Rozi{\`e}re, B., Goyal, N., Hambro, E., Azhar, F., et~al.: Llama: Open and
  efficient foundation language models. arXiv preprint arXiv:2302.13971  (2023)

\bibitem{wei2024proving}
Wei, J.T.Z., Wang, R.Y., Jia, R.: Proving membership in llm pretraining data
  via data watermarks. arXiv preprint arXiv:2402.10892  (2024)

\bibitem{yan2024protecting}
Yan, B., Li, K., Xu, M., Dong, Y., Zhang, Y., Ren, Z., Cheng, X.: On protecting
  the data privacy of large language models (llms): A survey. arXiv preprint
  arXiv:2403.05156  (2024)

\bibitem{yeom2018privacy}
Yeom, S., Giacomelli, I., Fredrikson, M., Jha, S.: Privacy risk in machine
  learning: Analyzing the connection to overfitting. In: 2018 IEEE 31st
  computer security foundations symposium (CSF). pp. 268--282. IEEE (2018)

\end{thebibliography}

\end{document}